\documentstyle[sprocl,epsf,12pt]{article}

\input{psfig.sty}

\def\be{\begin{equation}}
\def\ee{\end{equation}}
\def\bea{\begin{eqnarray}}
\def\eea{\end{eqnarray}}

\begin{document}


\title{The anomalous $\gamma \to \pi^+ \pi^0 \pi^-$ form factor and \\
the light--quark mass functions at low momenta}

\author{DUBRAVKO KLABU\v CAR}

\address{Physics Department, P.M.F.,
   Zagreb University, Bijeni\v{c}ka c. 32, Zagreb, Croatia}

\author{BOJAN BISTROVI\' C}

\address{ Center for Theoretical Physics, Laboratory for
Nuclear Science and Department of \\ Physics,
Massachusetts Institute of Technology, Cambridge, Massachusetts 02139  }

\maketitle\abstracts{
The $\gamma \to 3\pi$ form factor was calculated
in a simple--minded constituent model with a constant quark mass 
parameter, as well as in the Schwinger-Dyson approach.
The comparison of these and various other
theoretical results on this anomalous process,
as well as the scarce already available data
(hopefully to be supplemented by more accurate CEBAF data),
seem to favor Schwinger--Dyson modeling which would
yield relatively small low--momentum values of
the constituent (dynamically dressed) quark mass function.
}

\noindent 

\begin{figure}[t]
\centerline{\psfig{figure=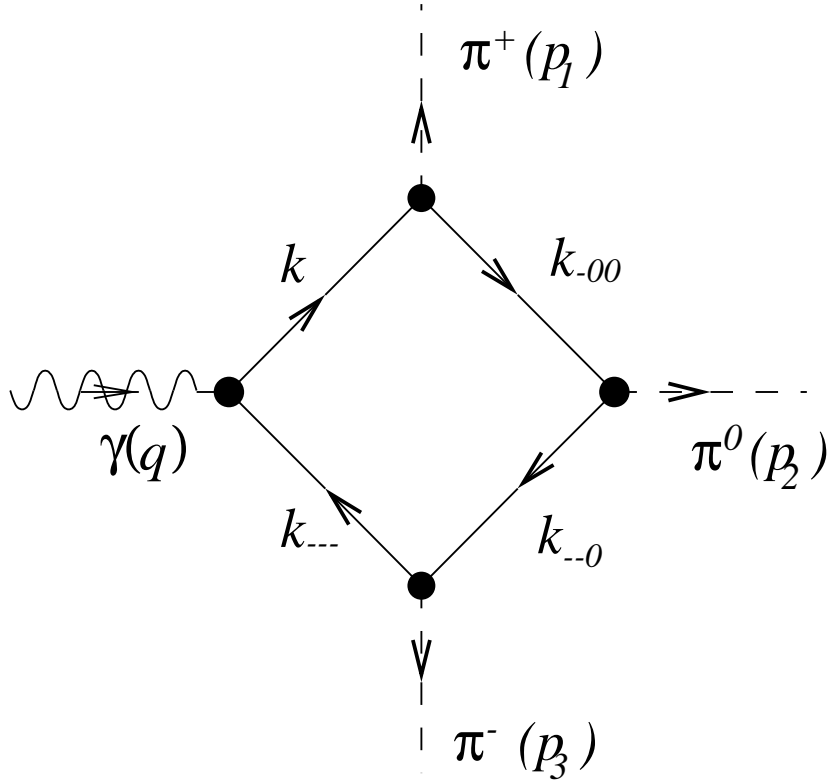,height=4.1in}}
\caption{One of the box diagrams for the process 
$\gamma(q)\to \pi^+(p_1) \pi^0(p_2) \pi^-(p_3)$. 
The other five are obtained from this one by the 
permutations of the vertices of the three different pions.
\label{fig:figure1}}
\end{figure}

The Abelian-anomalous $\pi^0\to\gamma\gamma$ amplitude is 
exactly \cite{Adler69,BellJackiw69}
$T_{\pi}^{2\gamma}(m_\pi = 0) = e^2 N_c /(12\pi^2 f_\pi)$
in the chiral and soft limit of pions of vanishing mass $m_\pi$.
On similarly fundamental grounds, the anomalous amplitude for the 
$\gamma(q)\to \pi^+(p_1) \pi^0(p_2) \pi^-(p_3)$ process, 
is predicted \cite{Ad+al71Te72Av+Z72}
to be 
\begin{equation}
F_\gamma^{3\pi}(0,0,0) \, = \, \frac{1}{e f_\pi^2} \, T_{\pi}^{2\gamma}(0) \, =
\, \frac{e N_c}{12 \pi^2 f_\pi^3} \, ,
\label{g3piAnomAmp}
\end{equation}
also in the chiral limit and at the soft point,
where the momenta of all three pions vanish:
$\{p_1,p_2,p_3\} = \{0,0,0\}$.
While the chiral and soft limit are an excellent approximation
for $\pi^0\to\gamma\gamma$, the already published \cite{Antipov+al87} 
and presently planned Primakoff experiments at CERN \cite{Moinester+al99},
as well as the current CEBAF measurement \cite{Miskimen+al94} of the
$\gamma(q)\to \pi^+(p_1) \pi^0(p_2) \pi^-(p_3)$,
involve values of energy and momentum transfer which are not
negligible compared to typical hadronic scales.
This gives a lot of motivation for theoretical predictions 
of the $\gamma\to 3 \pi$ amplitude for non--vanishing $\{p_1,p_2,p_3\}$,
{\it i.e.}, the form factor $F_\gamma^{3\pi}(p_{1},p_{2},p_{3})$.
We calculated it as the quark ``box"-amplitude (see Fig.~\ref{fig:figure1})
in the two related approaches \cite{BiKl99PRD,BiKl9912452} sketched below.

\begin{figure}[t]
\centerline{\psfig{figure=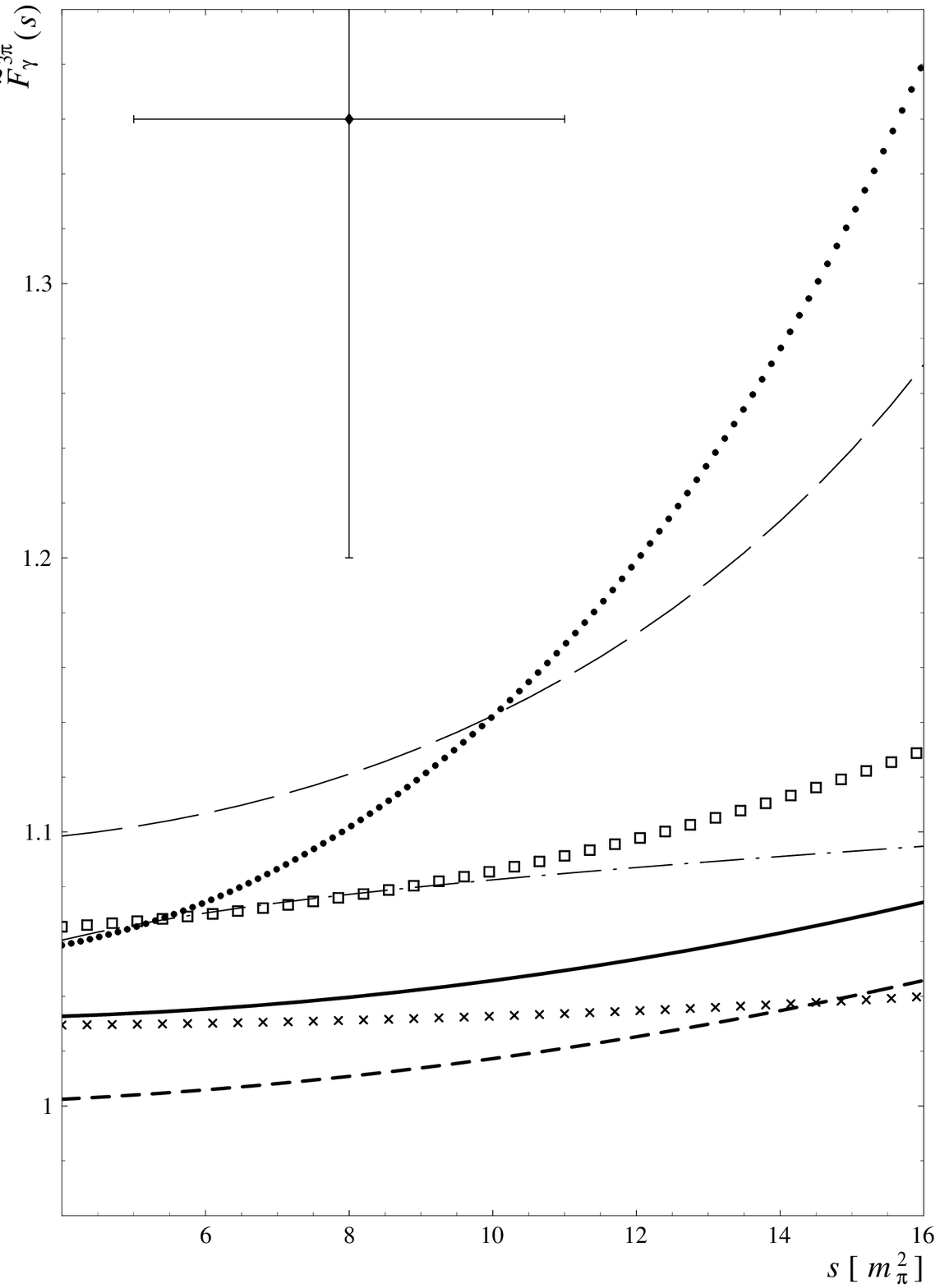,height=4.1in}}
\caption{Various predictions for the dependence of the normalized
$\gamma 3\pi$ form factor ${\widetilde{F}}^{3\pi}_\gamma$ on the
Mandelstam variable $s\equiv (p_1 + p_2)^2$. The kinematics is as
in the Serpukhov measurement (which provided$^4$ 
the shown data point): the photon and all three pions are
on shell, $q^2=0$ and $p_1^2=p_2^2=p_3^2=m_\pi^2$.  \label{fig:figure2}}
\end{figure}

In our Ref.~\cite{BiKl99PRD}, the intermediate fermion ``box" loop is 
the one of ``simple" constituent quarks with the constant quark mass 
parameter $M$.
The isospinor $\Psi = (u, d)^T$ of the light constituent quarks couple 
to the isovector pions $\pi^a$ through the pseudoscalar Yukawa coupling
$g\gamma_5\tau^a$. Its constant quark-pion coupling strength $g$
is related to the pion decay constant $f_{\pi} = 92.4$ MeV through the
quark-level Goldberger-Treiman (GT) relation $g/M = 1/f_{\pi}$.
The result of this calculation also corresponds to the form factor, 
in the lowest order in pion interactions, 
of the sigma-model and of the chiral quark model. 
In Ref. \cite{BiKl99PRD}, we give the analytic
expression for the form factor in terms of an expansion in the pion
momenta up to the order ${\cal O}(p^8)$ relative to the soft point result,
and also perform its exact numerical evaluation.
The latter predictions of this quark loop model \cite{BiKl99PRD} are given
[normalized to the soft-point amplitude (\ref{g3piAnomAmp})]
in Fig.~\ref{fig:figure2} by the long-dashed curve for $M=330 \, \rm{MeV}$,
by the line of empty boxes for $M=400 \, \rm{MeV}$, and by
the line of crosses for the large value $M=580 \, \rm{MeV}$.
Note that in the lowest order in pion interactions, they are also the 
form factors of the $\sigma$-model and of the chiral quark model.

Our second Ref. \cite{BiKl9912452} employs the Schwinger-Dyson (SD) 
approach \cite{Roberts0007054}, which is consistent both with the 
chiral symmetry constraints in the low-energy domain and with the 
perturbative QCD in the high-energy domain. In this approach, quarks 
in the fermion loop do not have free propagators with the simple-minded 
constant constituent mass $M$. Instead, the box loop amplitude is evaluated 
with the dressed quark propagator 
\begin{equation}
S(k)= \frac{1}{i \rlap{$k$}/ \,A(k^2) + m + B(k^2)}
       \equiv \frac{Z(k^2)}{i \, \rlap{$k$}/ \, + {\cal M}(k^2)}
\label{EuclS}
\end{equation}
containing the {\em momentum-dependent}, mostly dynamically generated 
quark mass function ${\cal M}(k^2)$ following from the SD solution for 
the dressed quark propagator (\ref{EuclS}). 
The explicit chiral symmetry breaking $m (\sim 2$ MeV 
in the present model choice \cite{AR96,BiKl9912452}) 
is two orders of magnitude smaller
than the quark mass function {\it at small momenta}, where 
it corresponds to the notion of the constituent quark mass.
Indeed, in Refs. \cite{AR96,BiKl9912452} as well as in the
model choice reviewed in our Ref. \cite{KeBiKl98}, 
${\cal M}(k^2\sim 0) \sim 300$ to 400 MeV.
On the other hand, since already the present--day SD modeling is
well--based \cite{Roberts0007054} on many aspects of QCD, such 
SD--generated ${\cal M}(k^2)$ should be close to the true QCD
quark mass function.

SD approach employs the Bethe--Salpeter (BS) bound--state 
pion--quark--antiquark vertex $\Gamma_{\pi^a}(k,p_{\pi^a})$
(here, in Fig.~\ref{fig:figure1}, instead of the aforementioned 
momentum--independent Yukawa coupling). The propagator (\ref{EuclS}) is 
consistent with the solution for the BS solution for 
$\Gamma_{\pi^a}(k,p_{\pi^a})$, and then, in this 
approach, the light pseudoscalar mesons are simultaneously the 
quark-antiquark bound states and the (quasi) Goldstone bosons of 
dynamical chiral symmetry breaking \cite{Roberts0007054}. 
Thanks to this, and also to 
carefully preserving the vector Ward-Takahashi identity in the 
quark-photon vertex, the {\it both} fundamental anomalous amplitudes 
$T_{\pi}^{2\gamma}(0)$ and $F_\gamma^{3\pi}(0,0,0)$ for the respective 
decays $\pi^0 \to \gamma \gamma$ and $\gamma \to \pi^+ \pi^0 \pi^-$, 
are evaluated analytically and exactly in the chiral limit and
the soft limit \cite{AR96}. (Note that reproducing these results 
even only roughly, let alone analytically, is otherwise quite problematic 
for bound-state approaches, as discussed in Ref. \cite{KeBiKl98}.)

In Fig.~\ref{fig:figure2}, the solid curve gives our 
$\gamma 3\pi$ form factor obtained 
in the SD approach for the empirical pion mass, $m_\pi = 138.5$ MeV,
while the dashed curve gives it in the chiral limit, $m_\pi = 0 = m$. 
To understand the relationship between the predictions of these two 
approaches, one should, besides the curves in Fig.~\ref{fig:figure2},
compare also the analytic expressions we derived for the 
form factors [esp. Eqs. (20)--(21) in Ref. \cite{BiKl9912452}
and analogous formulas in Ref. \cite{BiKl99PRD}]. 
This way, one can see, first, why the constant, 
momentum-independent term is smaller in the SD case, causing 
the downward shift of the SD form factors with respect to those 
in the constant constituent mass case. Second, this constant term 
in the both approaches diminishes with the increase of the 
pertinent mass scales, namely $M$ in the constant-mass case, 
and the scale which rules the SD--modeling and which is of course 
closely related to the resulting scale of the {\em dynamically generated} 
constituent mass ${\cal M}(k^2\sim 0)$. Finally, the 
momentum--dependent terms are similar in the both approaches;
notably, the coefficients of the momentum expansions (in powers
of $p_i \cdot p_j$) are similarly suppressed by powers of
their pertinent scales. This all implies a
transparent relationship between ${\cal M}(k^2)$ at small $k^2$
and the $\gamma 3\pi$ form factor, so that the accurate CEBAF data,
which hopefully are to appear soon \cite{Miskimen+al94}, 
should be able to constrain ${\cal M}(k^2)$ at small $k^2$,
and thus the whole infrared SD modeling. 
Admittedly, we used the Ball--Chiu Ansatz for the dressed quark--photon 
vertex, 
but this is adequate since Ref. \cite{MarisTandy9910033} found that for 
$-0.4 \, {\rm GeV}^2 < q^2 < 0.2 \, {\rm GeV}^2$, the true solution 
for the dressed vertex is approximated well by this Ansatz plus 
the vector--meson resonant contributions which however vanish 
in our case of the real photon, $q^2=0$.
Therefore, 
if the experimental form factor is measured with sufficient precision 
to judge the present SD model results definitely too low, it will
be a clear signal that the SD modeling should be reformulated
and refitted so that it is governed by a smaller mass scale and 
smaller values of ${\cal M}(k^2\sim 0)$.

The only already available data, the Serpukhov experimental 
point \cite{Antipov+al87} (shown in the upper left corner of 
Fig.~\ref{fig:figure2}), is higher than all theoretical predictions
and is probably an overestimate. However, the SD predictions are
farthest from it. Indeed, in the momentum interval shown in 
Fig.~\ref{fig:figure2}, the SD form factors are lower than those of 
other theoretical approaches (for reasonable values of their parameters) 
including vector meson dominance \cite{Rudaz84} (the dotted curve) and of 
chiral perturbation theory \cite{Holstein96} (the dash-dotted curve).
Therefore, even the present experimental and theoretical
knowledge indicates that the momentum--dependent mass function
in the SD model \cite{AR96} we adopted \cite{BiKl9912452},
may already be too large at small $k^2$, where its typical value for
light $u, d$ quarks is ${\cal M}(k^2\approx 0) \approx 360$ MeV.
Note that this value is, at present, probably the lowest in
the SD--modeling except for the model reviewed in Ref. \cite{KeBiKl98},
which has very similar ${\cal M}(k^2)$ at low $k^2$.  
(Some other very successful~\cite{Roberts0007054} SD models
obtain even higher values, ${\cal M}(k^2\approx 0)\approx 600$ MeV
and more,
which would lead to even lower $\gamma 3\pi$ transition form factors.)
It is thus desirable to reformulate SD phenomenology using
momentum--dependent mass functions which are smaller at low $k^2$.
This conclusion is in agreement with recent lattice QCD studies of 
the quark propagator which find \cite{SkullWilli0007028}
${\cal M}(k^2=0)=298\pm 8$ MeV (for $m=0$).

\vskip 3mm

\noindent {\bf Acknowledgment:}
D. Klabu\v car thanks the organizers, M. Rosina and B. Golli, for 
their hospitality and for the partial support which made possible
his participation at Mini-Workshop Bled 2000: FEW-QUARK PROBLEMS,
Bled, Slovenia, 8--15 July 2000.

\end{document}